\documentclass[showpacs,twocolumn,prl,aps]{revtex4}
\usepackage{graphicx}

\begin{document}

\title[Short title for running header]{Spontaneous quantum Hall effect in quarter-doped Hubbard model on honeycomb lattice and its possible realization in quarter-doped graphene system}
\author{Tao Li}
\affiliation{ Department of Physics, Renmin University of China,
Beijing 100872, P.R.China}
\date{\today}

\begin{abstract}
We show that as the result of the nesting property of the Fermi
surface, the quarter-doped Hubbard model on honeycomb lattice is
unstable with respect to the formation of a magnetic insulating
state with nonzero spin chirality for infinitesimally small strength
of electron correlation. The insulating state is found to be
topological nontrivial and to have a quantized Hall conductance of
$\sigma_{xy}=\frac{e^{2}}{h}$. We find the Fermi surface nesting is
robust for arbitrary value of next-nearest-neighbor hopping
integral. It is thus very possible that the quarter-doped graphene
system will realize such an exotic ground state. We also show that
the quarter-doped Hubbard model on honeycomb lattice is in exact
equivalence in the weak coupling limit with the 3/4-filled Hubbard
model on triangular lattice, in which similar effect is also
observed.
\end{abstract}
 \maketitle

Effects of electron correlation on honeycomb lattice have attracted
a lot of interest recently\cite{Meng,Xu,Wang,Ran,Li1}. The honeycomb
lattice has the smallest coordination number of 3 for a two
dimensional lattice and has Dirac-type dispersion at half filling.
It is a nontrivial task to understand the role of electron
correlation in such a background. As an example, it is recently
reported that the transition from the semimetal phase to the
antiferromagnetic ordered phase in the half-filled Hubbard model on
honeycomb lattice is realized in a two-step manner. An exotic spin
liquid state is found in a small but finite correlation range in
between the semimetal phase and the antiferromagnetic ordered
phase.\cite{Meng}.

However, in the two dimensional graphene sheet\cite{Geim,Neto},
which is thought to be the most natural realization of an electron
system with a honeycomb lattice, the correlation effect is believed
to be rather weak. Such a view is in general reasonable. However,
when the Fermi energy approaches the Van Hove singularity(VHS) of
the density of state, the importance of the correlation effect is
much enhanced. Especially, when the system has a perfectly nested
Fermi surface, an infinitesimally small interaction is enough to
induce dramatic correlation effect. Recently, large electron doping
on graphene system has been achieved in experiment through both
chemical doping and electrolytic gating\cite{ARPES,doping}. It is
also found by angle-resolved-photoemission measurement that the VHS
is much more extended than the free electron prediction.

The density of state of the free electron on honeycomb lattice (with
nearest-neighbor hopping) is shown in Fig.1. The VHS appears on both
sides of the half-filled background(at $3/4$ or $5/4$ band
filling\cite{filling}). At such special fillings, the system has a
perfectly nested Fermi surface. As an example, the single particle
energy on the Brillouin zone boundary is given by $\pm t$, where $t$
is the hopping integral between neighboring sites. It is important
to note that rather than one, there are three independent nesting
vectors: $\mathrm{Q}_{1}=\frac{1}{2}\mathbf{b}_{1}$,
$\mathrm{Q}_{2}=\frac{1}{2}\mathbf{b}_{2}$ and
$\mathrm{Q}_{3}=\frac{1}{2}(\mathbf{b}_{1}+\mathbf{b}_{2})$, where
$\mathbf{b}_{1}$ and $\mathbf{b}_{2}$ are the two elementary
reciprocal vectors of the system(see Fig.1). The exact same
situation also occurs on triangular lattice when the band is
$\frac{3}{4}$-filled. In that case, an interesting magnetic
insulating state with nonzero spin chirality and spontaneous quantum
Hall effect is predicted\cite{Martin,Batista}. In this paper, we
find that the exact same thing will also happen on the honeycomb
lattice. In fact, we can show that both systems share the exact same
low energy theory near the VHS, apart from an overall reduction of
the energy scale by a factor of 2. We also find that the nesting
property of the Fermi surface is robust for arbitrary value of the
next-nearest-neighbor hopping integral. We thus argue that the
quarter-doped graphene system has large chance to realize such an
exotic spontaneous quantum Hall state.

\begin{figure}[h!]
\includegraphics[width=8cm,angle=0]{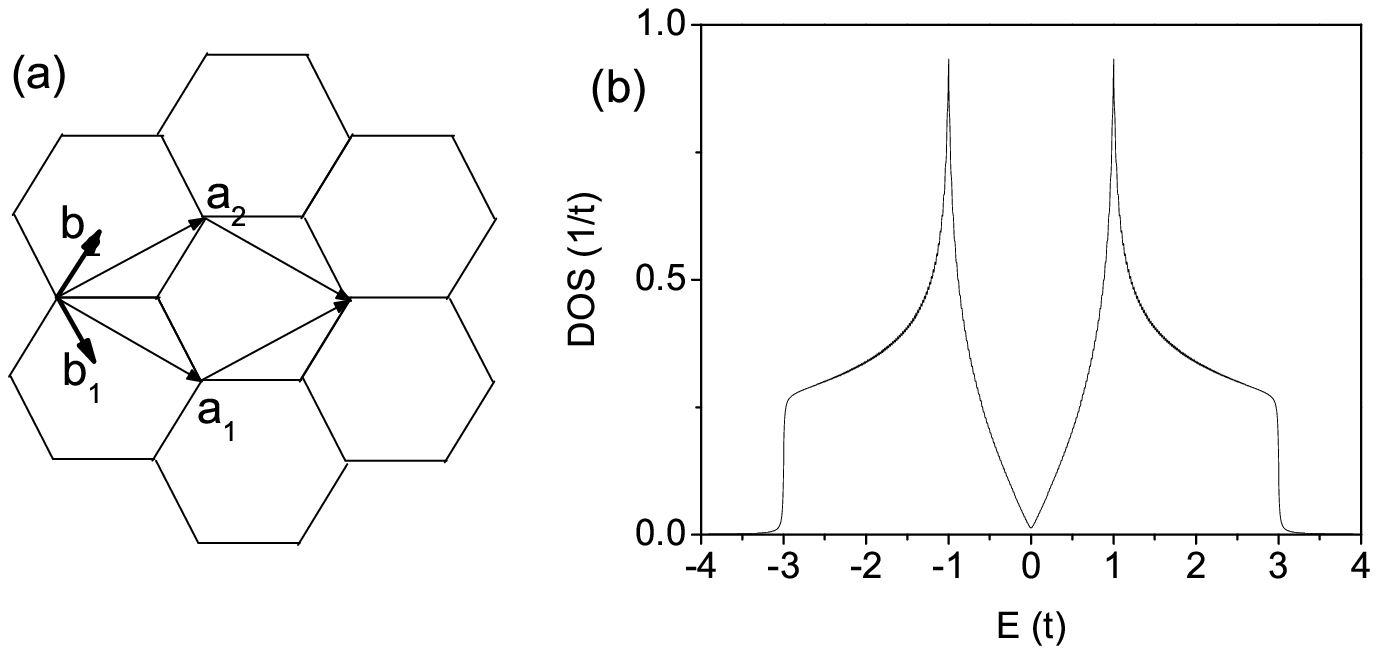}
\includegraphics[width=8cm,angle=0]{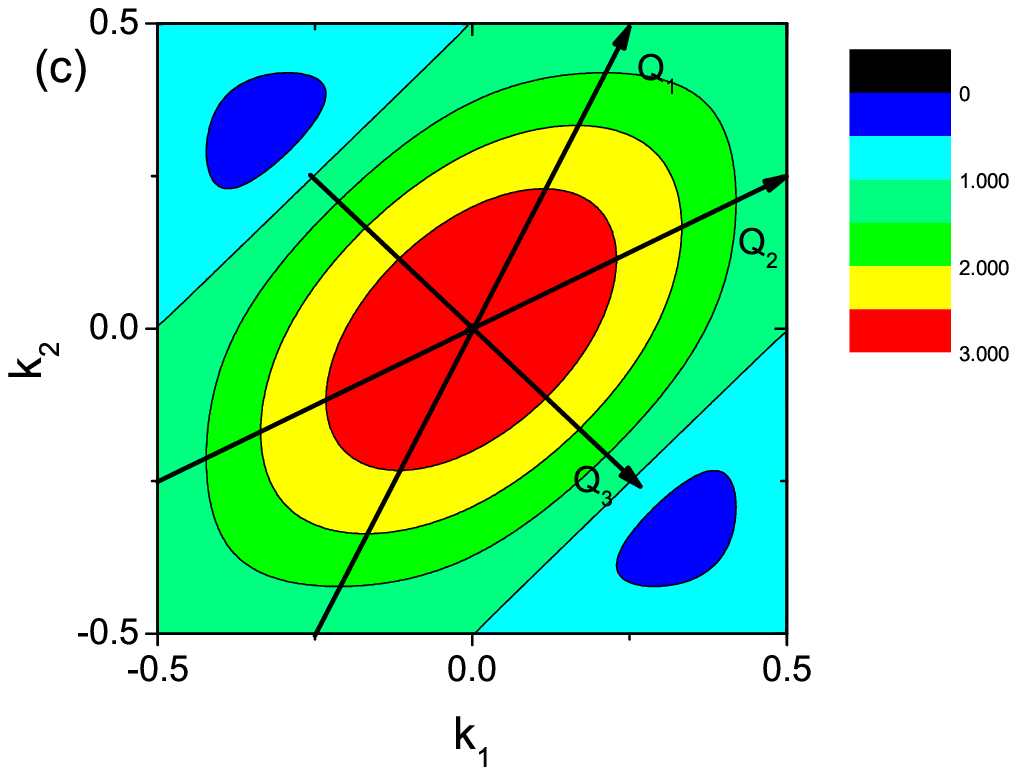}
\caption{The density of state and the Fermi surface of free electron
on honeycomb lattice with nearest-neighbor hopping. The VHS appears
at $E=\pm t$, which corresponds to electron filling of $3/4$ and
$5/4$ per unit cell. } \label{fig1}
\end{figure}

In a magnetic ordered background with nonzero spin chirality, the
motion of electron along a closed loop will experience a Berry
phase(equals to the solid angle subtended by the spins along the
loop), whose effect is indistinguishable from the Aharonov-Bohm
phase induced by external magnetic field. Such a quantum phase is
predicted to induce anomalous contribution to the Hall coefficient
in systems with non-coplanar magnetic
order\cite{Nagaosa,Taguchi,Raghu}. In particular, when the magnetic
order opens a full gap in the electron spectrum, such anomalous
contribution can only take quantized values.

For the $\frac{3}{4}$-filled Hubbard model on triangular lattice, as
the result of the Fermi surface nesting, a four-sublattice
non-coplanar magnetic order is established for infinitesimally small
value of electron correlation. In this state, the ordered moments in
the four sublattices point along the normals of the four surfaces of
a tetrahedron. The spins on each elementary triangular plaquette
subtend a solid angle of $\pi$. In such a non-coplanar state, the
electron spectrum opens a full gap and the Berry phase effect
induces a quantized Hall conductance of
$\sigma_{xy}=\frac{e^{2}}{h}$ at zero temperature. Since the order
is deduced from a weak coupling instability related to the nesting
property of the Fermi surface, it should be stable when the
correlation is not too strong.

Since the system has three independent nesting vectors at the VHS,
it has many different choices for developing a magnetic order. The
simplest choice of Bose condensation at a single nesting vector
results in a collinear magnetic state, while the non-coplanar
magnetic state with the tetrahedron ordering pattern can be viewed
as a state with Bose condensation on all the three nesting vectors
with the same strength\cite{Martin}. The tetrahedron ordering
pattern can be perfectly fitted into the honeycomb lattice as shown
in Fig.2. In such a state, each site is neighbored by three sites in
the other three magnetic sublattices. We note that the system can
also be viewed as being composed of two triangular sublattices, on
each of which a tetrahedron ordering pattern is established.

\begin{figure}[h!]
\includegraphics[width=7cm,angle=0]{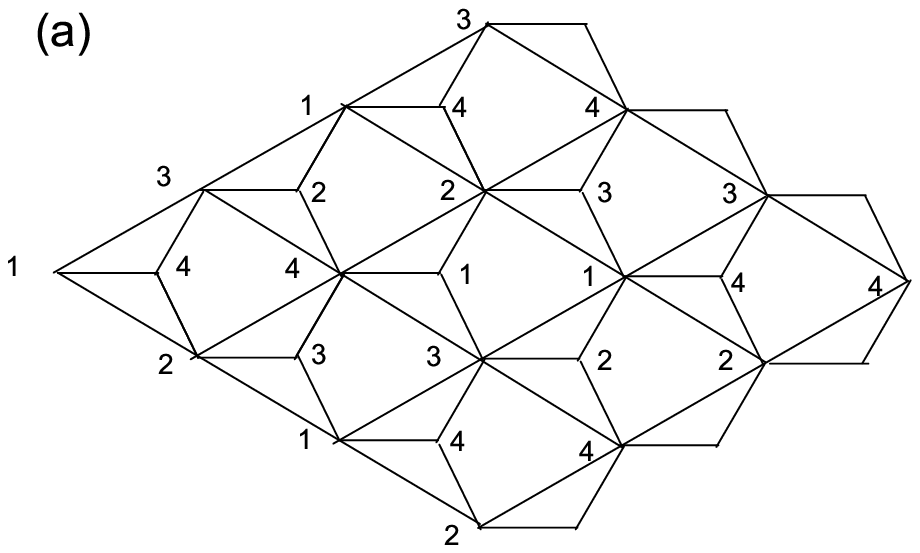}
\includegraphics[width=7cm,angle=0]{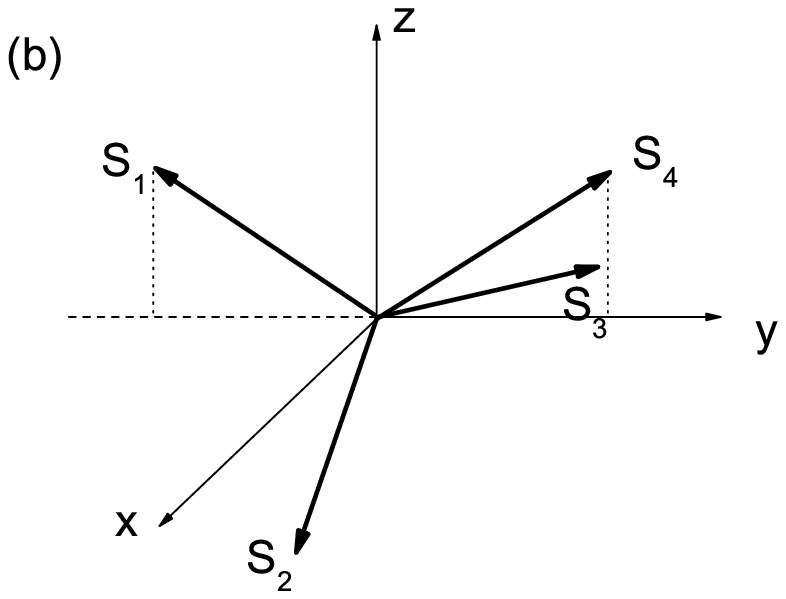}
\caption{Illustration of the tetrahedron magnetic ordering pattern
on honeycomb lattice. The directions of the ordered moments have
been chosen in such a way to make the mean field Hamiltonian as
simple as possible. } \label{fig2}
\end{figure}

To check the stability of the tetrahedron ordering pattern, we have
carried out unrestricted mean field search for the quarter-doped
Hubbard model on honeycomb lattice with $12\times12\times2$ lattice
sites. We have included both magnetic and charge order parameters in
the optimization. On the $12\times12\times2$ lattice, there are in
total 864 variational parameters for the spin density and 288
variational parameters for the charge density to be optimized. We
have used both the conjugate gradient method and the simulated
annealing method to find the minimum of the variational energy. Both
methods predict that the tetrahedron ordering pattern is the most
stable one and the charge distribution is always uniform. We now
present the mean field theory of this state.

In the mean field treatment, the Hubbard model on honeycomb lattice
has the form,
\begin{eqnarray}
H&=&-t\sum_{<i,j>,\sigma}(c^{\dagger}_{i,A,\sigma}c_{j,B,\sigma}+h.c.)\\\nonumber
&-&\frac{4U}{3}\sum_{i}<\mathrm{S}_{i,A}>\cdot
\mathrm{S}_{i,A}-\frac{4U}{3}\sum_{j}<\mathrm{S}_{j,B}>\cdot
\mathrm{S}_{j,B},
\end{eqnarray}
here $\mathrm{S}_{i,A}(\mathrm{S}_{j,B})$ denotes the spin operator
on site $i(j)$ of sublattice $A(B)$.

The ordered moments in the tetrahedron state can be written as
\begin{eqnarray}
<\mathrm{S}_{i,A}>=\frac{m}{\sqrt{3}}(\vec{\mathbf{e}}_{z}e^{i\mathbf{Q}_{3}\cdot
\mathrm{R}_{i}}+\vec{\mathbf{e}}_{x}e^{i\mathrm{Q}_{1}\cdot
\mathrm{R}_{i}}+ \vec{\mathbf{e}}_{y}e^{i\mathrm{Q}_{2}\cdot
\mathrm{R}_{i}} )
\\\nonumber
<\mathrm{S}_{j,B}>=\frac{m}{\sqrt{3}}(\vec{\mathbf{e}}_{z}e^{i\mathrm{Q}_{3}\cdot
\mathrm{R}_{j}}-\vec{\mathbf{e}}_{x}e^{i\mathrm{Q}_{1}\cdot
\mathrm{R}_{j}}- \vec{\mathbf{e}}_{y}e^{i\mathrm{Q}_{2}\cdot
\mathrm{R}_{j}} ).
\end{eqnarray}
With Eq.(2), the mean field Hamiltonian in the momentum space takes
the form
\begin{equation}
H=\sum_{\mathrm{k}\in
MBZ}\psi_{\mathrm{I},\mathrm{k}}^{\dagger}(\Gamma_{\mathrm{k}}-\beta\mathrm{M})\psi_{\mathrm{I},\mathrm{k}}+\psi_{\mathrm{II},\mathrm{k}}^{\dagger}(\Gamma_{\mathrm{k}}-\beta\mathrm{M})\psi_{\mathrm{II},\mathrm{k}},
\end{equation}
in which
 $\psi_{\mathrm{I},\mathrm{k}}^{\dagger}=(c^{A\dagger}_{\mathrm{k}\uparrow},c^{B\dagger}_{\mathrm{k}\uparrow},c^{A\dagger}_{\mathrm{k+Q_{1}}\downarrow},c^{B\dagger}_{\mathrm{k+Q_{1}}\downarrow},c^{A\dagger}_{\mathrm{k+Q_{2}}\downarrow},c^{B\dagger}_{\mathrm{k+Q_{2}}\downarrow},\\c^{A\dagger}_{\mathrm{k+Q_{3}}\uparrow},c^{B\dagger}_{\mathrm{k+Q_{3}}\uparrow})$
,
 $\psi_{\mathrm{II},\mathrm{k}}^{\dagger}=(c^{A\dagger}_{\mathrm{k}\downarrow},c^{B\dagger}_{\mathrm{k}\downarrow},c^{A\dagger}_{\mathrm{k+Q_{1}}\uparrow},c^{B\dagger}_{\mathrm{k+Q_{1}}\uparrow},\\-c^{A\dagger}_{\mathrm{k+Q_{2}}\uparrow},-c^{B\dagger}_{\mathrm{k+Q_{2}}\uparrow},-c^{A\dagger}_{\mathrm{k+Q_{3}}\downarrow},-c^{B\dagger}_{\mathrm{k+Q_{3}}\downarrow})$.
The spectrum is thus explicitly two-fold degenerate. $\Gamma_{k}$
and $\mathrm{M}$ are $8\times8$ Hermitian matrices and are given by
\begin{eqnarray}
\Gamma_{\mathrm{k}}=\left(\begin{array}{cccc}
                      \gamma_{\mathrm{k}} & 0 & 0 & 0 \\
                      0 & \gamma_{\mathrm{k+Q_{1}}} & 0 & 0 \\
                      0 & 0 & \gamma_{\mathrm{k+Q_{2}}} & 0 \\
                      0 & 0 & 0 & \gamma_{\mathrm{k+Q_{3}}}
                    \end{array}\right)\nonumber
\end{eqnarray}
and
\begin{eqnarray}
\mathrm{M}=\left(\begin{array}{cccc}
                      0 & \sigma_{3} & -i\sigma_{3} & I \\
                      \sigma_{3} & 0 & -I & i\sigma_{3} \\
                      i\sigma_{3} & -I & 0 & \sigma_{3} \\
                      I & -i\sigma_{3} & \sigma_{3} & 0
                    \end{array}\right),\nonumber
\end{eqnarray}
in which $\gamma_{\mathrm{k}}=-\left(\begin{array}{cc}
                                \mu & g_{\mathrm{k}} \\
                                g^{*}_{\mathrm{k}} & \mu
                              \end{array}\right)
$, $g_{\mathrm{k}}=1+e^{-i2\pi k_{1}}+e^{-i2\pi k_{2}}$,
$\sigma=\left(\begin{array}{cc}
                                1 & 0 \\
                                0 & -1
                              \end{array}\right)$, $I$ is the
                              $2\times2$ identity matrix.
                              $\beta=\frac{2Um}{3\sqrt{3}}$ and $\mu$ is the
                              chemical potential to be determined by filling
                              concentration. In the summation, the
                              momentum is restricted to the magnetic
                              Brillouin zone given by
                              $-\frac{1}{2}\leq k_{1},k_{2}\leq\frac{1}{2}$(here we have used the convention for momentum in which $\mathbf{k}=k_{1}\mathbf{b}_{1}+k_{2}\mathbf{b}_{2}$).

The self-consistent equation for the order parameter $m$ is given by
\begin{equation}
m=\frac{1}{4\sqrt{3}N}\sum_{\mathrm{k}\in
MBZ}\left[\langle\psi_{\mathrm{I},\mathrm{k}}^{\dagger}\mathrm{M}\psi_{\mathrm{I},\mathrm{k}}\rangle+\langle\psi_{\mathrm{II},\mathrm{k}}^{\dagger}\mathrm{M}\psi_{\mathrm{II},\mathrm{k}}
\rangle\right].
\end{equation}
We have solved this equation as a function of $U/t$ at zero
temperature for the quarter-doped system. The result is shown in
Fig.3. The order parameter $m$ is seen to increase with $U/t$ from
$U/t=0$. This is the result of the nesting property of the Fermi
surface. The tetrahedron order is thus a weak coupling instability
of the system.
\begin{figure}[h!]
\includegraphics[width=6cm,angle=0]{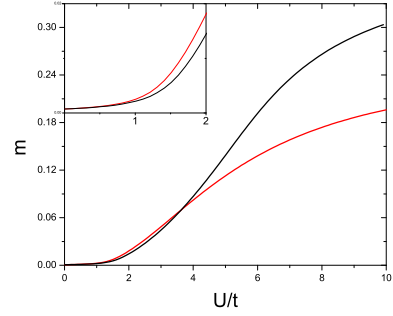}
\caption{The order parameter as a function of $U/t$ at zero
temperature. The black line gives the result of the quarter-filled
honeycomb lattice, the red line gives the result of the 3/4-filled
triangular lattice for a comparison. The inset shows the comparison
in the weak coupling limit, in which the two lattices show the exact
same behavior. } \label{fig3}
\end{figure}

The tetrahedron state has a nonzero spin chirality, $i.e.$,
$\langle\mathrm{S}_{i}\cdot(\mathrm{S}_{j}\times\mathrm{S}_{k})\rangle\neq0$
for spins on neighboring sites. Electron moving on such a background
will experience a nonzero Berry phase, which will induce an
anomalous contribution to the Hall response of the electron system.
Since the state is fully gapped, such an anomalous contribution must
take quantized values of the form $\frac{ne^{2}}{h}$, where $n$ is
the total Chern-number of all the occupied bands. Now we show that
$\sigma_{xy}=\frac{e^{2}}{h}$ at zero temperature in this state.

The Kubo formula for the Hall conductance $\sigma_{xy}$ is given by
\begin{equation}
\sigma_{xy}=\frac{e^{2}}{h}\frac{2\pi}{iS}\sum_{\mathrm{k},n,m}\frac{j^{x}_{n,m}(\mathrm{k})j^{y}_{m,n}(\mathrm{k})}{(\epsilon_{n}(\mathrm{k})-\epsilon_{m}(\mathrm{k}))^{2}}(f(\epsilon_{n}(\mathrm{k}))-f(\epsilon_{m}(\mathrm{k}))),
\end{equation}
in which $S$ is the area of the system. $j^{x}_{n,m}(\mathrm{k})$
and $j^{y}_{n,m}(\mathrm{k})$ are the matrix elements of the current
operators between the n-th and the m-th eigenstate of the band
Hamiltonian at momentum $\mathrm{k}$. $f(\epsilon_{m}(\mathrm{k}))$
is the Fermi distribution function. The current operator $j^{x}$ is
given by
\begin{eqnarray}
j^{x}=\sum_{\mathrm{k}\in
MBZ}\psi_{\mathrm{I},\mathrm{k}}^{\dagger}\mathrm{C}^{x}_{\mathrm{k}}\psi_{\mathrm{I},\mathrm{k}}+\psi_{\mathrm{II},\mathrm{k}}^{\dagger}\mathrm{C}^{x}_{\mathrm{k}}\psi_{\mathrm{II},\mathrm{k}}.
\end{eqnarray}
$j^{y}$ is given by a similar expression with
$\mathrm{C}^{x}_{\mathrm{k}}$ replaced by
$\mathrm{C}^{y}_{\mathrm{k}}$. Here $\mathrm{C}^{x,y}_{\mathrm{k}}$
are $8\times8$ matrices and are given by
\begin{eqnarray}
\mathrm{C}^{x,y}_{\mathrm{k}}=\left(\begin{array}{cccc}
                      c^{x,y}_{\mathrm{k}} & 0 & 0 & 0 \\
                      0 & c^{x,y}_{\mathrm{k+Q_{1}}} & 0 & 0 \\
                      0 & 0 & c^{x,y}_{\mathrm{k+Q_{2}}} & 0 \\
                      0 & 0 & 0 & c^{x,y}_{\mathrm{k+Q_{3}}}
                    \end{array}\right),\nonumber
\end{eqnarray}
in which $c^{x,y}_{\mathrm{k}}=\left(\begin{array}{cc}
                                0 & v^{x,y}_{\mathrm{k}} \\
                                v^{*x,y}_{\mathrm{k}} & 0
                              \end{array}
\right)$, $v^{x}_{\mathrm{k}}=it-\frac{it}{2}(e^{-i2\pi
k_{1}}+e^{-i2\pi k_{2}})$,
$v^{y}_{\mathrm{k}}=\frac{it\sqrt{3}}{2}(e^{-i2\pi k_{1}}-e^{-i2\pi
k_{2}})$.

To calculate $\sigma_{xy}$, we first solve the mean field equation
at finite temperature. The order parameter as a function of
temperature at fixed doping and $\frac{U}{t}=4$ is shown in Fig.4.
The mean field critical temperature is found to be
$T_{c}\approx0.09t$ for this set of parameters. The Hall
conductivity is also shown in Fig.4. $\sigma_{xy}$ is found to be
zero above $T_{c}$ and begins to increase below $T_{c}$, following
the exact same trend as the order parameter. It finally saturates to
the quantized value of $\sigma_{xy}=\frac{e^{2}}{h}$ at zero
temperature.
\begin{figure}[h!]
\includegraphics[width=7cm,angle=0]{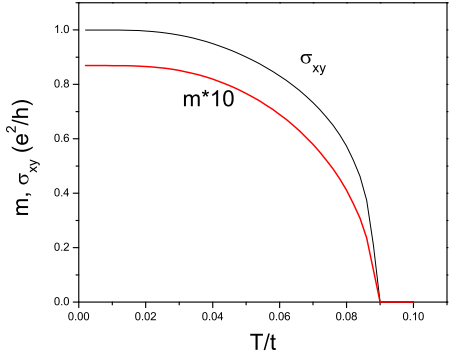}
\caption{The order parameter and the Hall conductance as functions
of temperature at $U/t=4$. Note the close similarity of these
results with the results on triangular lattice\cite{Martin}.}
\label{fig4}
\end{figure}

The results obtained here are very similar to the results for the
$3/4$-filled Hubbard model on triangular lattice\cite{Martin}. For
example, the critical temperatures on both lattices are around
$0.09t$ for $\frac{U}{t}=4$. To understand this close similarity, we
note that the honeycomb lattice is actually composed of two
interpenetrating triangular lattices, on each of which a tetrahedron
ordering pattern is established. Actually, we can show that the
quarter-doped Hubbard model on honeycomb lattice and $3/4$-filled
Hubbard model on triangular lattice are described by the same theory
in the low energy limit, except for a rescaling of the overall
energy scale by a factor of 2. First, we note the dispersion of the
honeycomb lattice has the form
\begin{eqnarray}
E^{h}_{\mathrm{k}}=t|g(\mathrm{k})|=t\sqrt{3+2(\cos(k_{1})+\cos(k_{2})+\cos
(k_{1}-k_{2}))} \nonumber,
\end{eqnarray}
while the dispersion of the triangular lattice has the form
\begin{eqnarray}
E^{t}_{\mathrm{k}}=-2t(\cos(k_{1})+\cos(k_{2})+\cos
(k_{1}-k_{2}))\nonumber.
\end{eqnarray}
It can then be shown that both lattices share the same dispersion
relation near the VHS(where $E^{t}_{\mathrm{k}}=-2t$) apart from a
factor of 2 rescaling of the energy scale. At the same time, it can
be shown that the Hubbard interaction term on both lattices can be
rescaled into each other in the low energy limit and the rescaling
factor is also 2. In the low energy limit, the Hubbard interaction
$H_{U}=U\sum_{i}n_{i,\uparrow}n_{i,\downarrow}$ on triangular
lattice is given by
\begin{equation}
H_{U}\sim
U\sum_{\mathrm{k}_{i},\mathrm{k}'_{i},i}c^{\dagger}_{\mathrm{k}_{i},\uparrow}c_{\mathrm{k}'_{i}+\mathrm{Q}_{i},\uparrow}c^{\dagger}_{\mathrm{k}'_{i},\downarrow}c_{\mathrm{k}_{i}+\mathrm{Q}_{i},\downarrow},
\end{equation}
where the sum over $i$ denotes the sum over the three patches of
nested Fermi surfaces. $\mathrm{k}_{i},\mathrm{k}'_{i}$ are the
momentums on the $i-$th patch of nested Fermi surface and
$\mathrm{Q}_{i}$ denotes the nesting vector on this patch of Fermi
surface. On honeycomb lattice, the Hubbard interaction takes the
form of
$H_{U}=U\sum_{i}(n_{iA,\uparrow}n_{iA,\downarrow}+n_{iB,\uparrow}n_{iB,\downarrow})$.
In the low energy limit, it reduces to
\begin{eqnarray}
H_{U}\sim
U\sum_{\mathrm{k}_{i},\mathrm{k}'_{i},i}c^{\dagger}_{A,\mathrm{k}_{i},\uparrow}c_{A,\mathrm{k}'_{i}+\mathrm{Q}_{i},\uparrow}c^{\dagger}_{A,\mathrm{k}'_{i},\downarrow}c_{A,\mathrm{k}_{i}+\mathrm{Q}_{i},\downarrow},\nonumber\\
+U\sum_{\mathrm{k}_{i},\mathrm{k}'_{i},i}c^{\dagger}_{B,\mathrm{k}_{i},\uparrow}c_{B,\mathrm{k}'_{i}+\mathrm{Q}_{i},\uparrow}c^{\dagger}_{B,\mathrm{k}'_{i},\downarrow}c_{B,\mathrm{k}_{i}+\mathrm{Q}_{i},\downarrow}.
\end{eqnarray}
The free electron system on honeycomb lattice forms two bands with
eigen-energy $E_{\pm,\mathrm{k}}=\pm t|g(\mathrm{k})|$. If we denote
the operators of the two bands as $c_{+,\mathrm{k}}$ and
$c_{-,\mathrm{k}}$, we have
\begin{eqnarray}
\left(\begin{array}{c}
                      c_{+,\mathrm{k}} \\
                      c_{-,\mathrm{k}}
                    \end{array}\right)
                    =\frac{1}{\sqrt{2}}\left(\begin{array}{cc}
                      1 & e^{i\phi_{\mathrm{k}}} \\
                      -e^{-i\phi_{\mathrm{k}}} & 1
                      \end{array}\right)\left(\begin{array}{c}
                      c_{A,\mathrm{k}} \\
                      c_{B,\mathrm{k}}
                    \end{array}\right),
\end{eqnarray}
in which $\phi_{\mathrm{k}}$ is the phase of $g(\mathrm{k})$. Noting
that the nested Fermi surface patches are given by
$k_{1}=\frac{1}{2}$, $k_{2}=\frac{1}{2}$ and
$k_{1}-k_{2}=\frac{1}{2}$, it is straightforward to show that
$\phi_{\mathrm{k}_{i}}-\phi_{\mathrm{k}_{i}+\mathrm{Q}_{i}}=\pi$,
$\pi$, and $0$ on the three patches. Thus, when the Hubbard
interaction term is projected into the subspace of the $E_{+}$(or
$E_{-}$) band, the phase of its matrix element will be exact zero.
Collecting all contributions, we find that in the low energy limit
the Hubbard interaction takes the form
\begin{eqnarray}
H_{U}\sim
\frac{U}{2}\sum_{\mathrm{k}_{i},\mathrm{k}'_{i},i}c^{\dagger}_{+,\mathrm{k}_{i},\uparrow}c_{+,\mathrm{k}'_{i}+\mathrm{Q}_{i},\uparrow}c^{\dagger}_{+,\mathrm{k}'_{i},\downarrow}c_{+,\mathrm{k}_{i}+\mathrm{Q}_{i},\downarrow},\nonumber\\
+\frac{U}{2}\sum_{\mathrm{k}_{i},\mathrm{k}'_{i},i}c^{\dagger}_{-,\mathrm{k}_{i},\uparrow}c_{-,\mathrm{k}'_{i}+\mathrm{Q}_{i},\uparrow}c^{\dagger}_{-,\mathrm{k}'_{i},\downarrow}c_{-,\mathrm{k}_{i}+\mathrm{Q}_{i},\downarrow}.
\end{eqnarray}
Thus in the low energy limit, the quarter-doped Hubbard model on
honeycomb lattice can be rescaled into the $3/4$-filled Hubbard
model on triangular lattice with a reduction of overall energy scale
by a factor of 2. This explains the close similarity between the
results obtained on both lattices. Note that this equivalence only
holds in the weak coupling limit(see Fig.3).

Although the calculation is done at the mean field level, our
conclusion that the ground state of the system is a topological
insulator with a quantized Hall conductance should be robust for
sufficiently small $\frac{U}{t}$, since the magnetic ordering is
induced by the Fermi surface nesting. The situation at finite
temperature is more subtle, since the Wagner-Mermin theorem
prohibits spontaneous breaking of continuous symmetry at finite
temperature in two dimensional systems. However, as the ordered
state also breaks the discrete time reversal symmetry, possibility
still exists that the spin chirality may survive even without
ordered moment\cite{Domenge}. It is interesting to see if the spin
chirality can survive at finite temperature, when the ordered moment
is already washed out by thermal fluctuation.

Finally, we turn to the implications of our results on the physics
of graphene. First, we note that the dispersion of the graphene
system is not perfectly particle-hole symmetric. Uncertainty remains
in the strength of longer ranger hopping integrals. The
next-neighbor hopping integral is estimated to be in the range of
$0.02t\leq t'\leq0.2t$\cite{Neto}. A small but finite
next-next-neighbor hopping integral can also exist\cite{nnn}.
However, we want to emphasis that the nesting property of the
quarter-doped system is robust against the introduction of
next-neighbor hopping integral. The perfect nesting will only be
destroyed by the next-next-neighbor hopping, which is much smaller
than the nearest and next-nearest neighboring hopping integrals. We
thus believe that our result should be applicable to the graphene
system.

To be more specific, we have done the mean field calculation for
nonzero $t'/t$. The order parameter at zero temperature as a
function of $U/t$ for $t'/t=0.2$ and $t'/t=-0.2$ are shown in Fig.5,
in which they are compared with the result for $t'/t=0$. It is seen
that the tetrahedron magnetic order still develops at
infinitesimally small value of interaction, no matter what is the
sign of $t'$. The sign of $t'$ does play a role in reducing or
enhancing the magnitude of the order parameter. On the honeycomb
lattice, the sign of $t'$ will change under the particle-hole
transformation. For example, electron doping in a system with
positive $t'$ is equivalent to hole doping in a system with negative
$t'$. Thus both signs of $t'$ are meaningful for the graphene
system.
\begin{figure}[h!]
\includegraphics[width=7cm,angle=0]{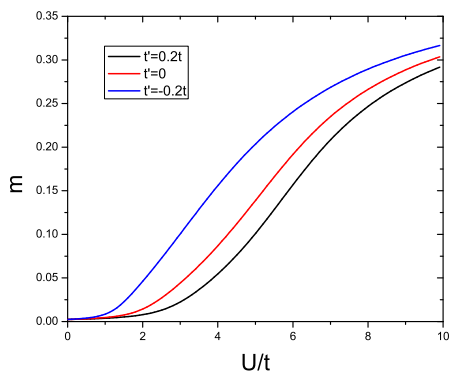}
\caption{The order parameter as a function of $U/t$ for $t'/t=0.2,0,
$ and $-0.2$. } \label{fig5}
\end{figure}

After the completion of this work\cite{Li2}, we noticed many related
recent researches in which other ordering patterns are
proposed\cite{Makogon,Castro,Chubukov,Murthy,Kiese,Lee}. Especially,
a chiral d-wave superconducting order is suggested for the ground
state of the quarter-doped system by RG
analysis\cite{Chubukov,Kiese}. However, more recent functional RG
study and variational calculation show that for the quarter-doped
system, the tetrahedron magnetic ordering pattern proposed in this
paper enjoys a much larger condensation energy than the chiral
d-wave superconducting state\cite{Lee}.

We conclude that the quarter-doped Hubbard model on honeycomb
lattice has weak coupling instability to the formation of a magnetic
insulating state with nonzero spin chirality and quantized Hall
conductance as a result of the nesting property of the Fermi
surface. We find the nesting property is robust against the
introduction of next-nearest-neighboring hopping terms. We thus
believe that such an exotic state should be observable in
quarter-doped graphene system. Such an exotic state can also act as
the parent phase of even more exotic phases if we move slightly away
from the VHS.

This work is supported by NSFC Grant No. 10774187 and National Basic
Research Program of China No. 2007CB925001 and No. 2010CB923004.


\begin{thebibliography}{9}
\bibitem{filling}Here the filling fraction is defined as the number
of electron per unit cell for each spin. The undoped graphene system
thus has a filling fraction of 1. Note this definition is different
from that used in \cite{Chubukov}, in which the undoped system has a
filling fraction of 0.5.
\end{thebibliography}
\end{document}